\documentclass[conference]{IEEEtran}
\IEEEoverridecommandlockouts

\usepackage{amsmath,amssymb,amsfonts}
\usepackage{graphicx}
\usepackage{booktabs}
\usepackage{multirow}
\usepackage{textcomp}
\usepackage{cite}

\begin{document}

\title{Enabling Emergency Communication via Semantic Radar-Centric ISAC}

\author{\IEEEauthorblockN{
Mohaimin Al Barat\IEEEauthorrefmark{1}, 
Chaoyu Zhang\IEEEauthorrefmark{1}, 
Hexuan Yu\IEEEauthorrefmark{1}, 
Kai Zeng\IEEEauthorrefmark{2}, 
Y. Thomas Hou\IEEEauthorrefmark{1}, 
and Wenjing Lou\IEEEauthorrefmark{1}
}
\IEEEauthorblockA{\IEEEauthorrefmark{1}Virginia Tech, USA}
\IEEEauthorblockA{\IEEEauthorrefmark{2}George Mason University, USA}
}

\maketitle

\begin{abstract}
Combat aircraft and other modern military platforms field two co-located RF assets: a radar for sensing and a dedicated radio for communication. In contested battlefield environments, however, the dedicated communication path is highly likely to be targeted and disrupted by jamming, blockage, or physical damage, while the radar remains fully operational. A natural fallback is to repurpose the surviving radar aperture as an emergency communication channel. The difficulty is that radar is purposely engineered for sensing rather than communication, so loading any information onto its waveform perturbs the beampattern and inevitably erodes sensing performance: the more bits we push, the more sensing we lose. We identify the core problem as \emph{enabling emergency communication via a radar-centric integrated sensing and communication (ISAC) link while preserving the radar's primary sensing function}. Therefore, we propose a semantic radar-centric ISAC framework that pairs a learned vector-quantized semantic encoder, which compresses task-relevant content into a few tens to a few hundred bits, with a wideband phased-MIMO radar restricted to a conservative spatial-spectral configuration subset that respects a bounded sensing-degradation budget. Experiments on SAR target classification and text classification show that our framework supports reliable emergency communication for downstream tasks while keeping the radar beampattern almost unchanged.
\end{abstract}


\section{Introduction}
\label{sec:intro}

Modern military platforms, such as combat aircraft, surface vessels, ground vehicles, and unmanned systems, typically carry co-located RF assets: a radar for surveillance, tracking, and fire control, and a separate tactical datalink or beyond-line-of-sight terminal for command, control, and situational awareness. In contested battlefield environments, the dedicated communication path is highly likely to be targeted and disrupted, leaving the platform without its primary link. Hostile electronic attack can suppress tactical links through barrage and spot jamming, spoofing, and protocol-level disruption; physical-layer exploits include spatially selective interference~\cite{mackensen2025ris_ndss}, commodity low-cost jammers~\cite{yang2024uwbad_ccs}, learning-assisted waveform attacks~\cite{schutz2024linjam_icc}, and targeted disruption of channel access~\cite{stegmann2024prach_globecomw}. Decades of electronic-warfare and jamming studies~\cite{pirayesh2022jamming_cst} reinforce the same operational picture: even brief jamming, antenna blockage, or battle damage can sever coordination with higher headquarters and wingmen, while the onboard radar stack typically keeps operating for local sensing.

When the dedicated datalink is down but the radar still works, the radar aperture is the obvious surviving RF resource for emergency information transfer. Joint radar and communication design has long studied how to embed bits into a radar emission~\cite{liu2020jrc_tcom}: representative mechanisms convey information through sidelobe modulation~\cite{hassanien2016sidelobe_tsp} or index modulation across spatial and spectral resources~\cite{huang2020majorcom_tsp}, and recent systems even support two-way radar-backscatter communication on commercial FMCW radars~\cite{okubo2024biscatter_sigcomm}. These efforts show that radar can, in principle, carry bits.

The fundamental difficulty is that radar is \emph{purposely} engineered for sensing rather than communication. Within a fixed radar aperture, communication and sensing compete for the same spatial, spectral, and waveform degrees of freedom: stronger modulation, denser symbol packing, or more aggressive waveform perturbation deforms the beampattern, raises sidelobes, and distorts the matched-filter response, all of which degrade sub-task performance~\cite{liu2018waveform_tsp,kang2019spatspect_taes,blunt2010intrapulse_taes, zhang2023mindfl, zhang2025enabling}. In short, \emph{the more communication we load onto the radar, the more sensing we lose}. We therefore target a central question: \emph{how can we enable emergency communication via a radar-centric ISAC link while preserving the radar's primary sensing function?}

Our key insight is that, in emergency settings, the platform rarely needs to transmit raw sensor data; it needs to convey a small amount of task-relevant information such as an object label, a relative geometry, or a compact perception state. If the message can be made small enough, the radar only has to carry a few tens to a few hundred bits, which can be accommodated by transmission configurations chosen to satisfy conservative sensing-preservation constraints. This naturally pairs radar-centric ISAC with task-oriented semantic communication, where a learned encoder transmits compact task-sufficient features rather than raw data~\cite{xie2021deepsc_tsp,shao2022ib_jsac,vqvae2017nips}. Joint sensing and task-oriented communication has recently emerged as an active ISAC direction~\cite{sagduyu2024joint}, but prior work still assumes a conventional communication link rather than a sensing-preserving radar-centric emergency channel. We therefore propose \textbf{semantic radar-centric ISAC}: a learned vector-quantized semantic encoder compresses the payload into a compact discrete representation, and a wideband phased-MIMO radar is restricted to a conservative spatial-spectral configuration subset whose beamforming vectors satisfy mainlobe, sidelobe, and user-gain constraints. The two designs are coupled by a single bandwidth-budget inequality, so that the semantic bit budget is matched to the radar-safe rate under a bounded sensing-degradation constraint, rather than being designed in isolation.

\section{Related Work}
\label{sec:related}

\textbf{Integrated sensing and communication.}
ISAC and dual-function radar-communication (DFRC) share one RF front end between sensing and information transfer~\cite{liu2022isac_jsac,liu2022fundamentals_cst,liu2020jrc_tcom}: communication-centric designs overlay sensing on a communication waveform, while radar-centric designs overlay communication on a radar waveform. With the dedicated radio denied but the radar operational, we adopt the radar-centric side. 

\textbf{Radar-centric communication.}
Radar-centric ISAC embeds bits into the radar emission. MIMO precoding characterizes the sensing--communication tradeoff~\cite{liu2018waveform_tsp}, sidelobe modulation carries information outside the main beam~\cite{hassanien2016sidelobe_tsp}, and symbol-level precoding~\cite{liu2022slp_jsac}, joint secure designs~\cite{dong2023joint}, and index modulation~\cite{huang2020majorcom_tsp,jin2024reconfigurable} extend this line. When sensing is primary, communication must fit a sensing-safe envelope: intrapulse embedding preserves the ambiguity function~\cite{blunt2010intrapulse_taes}, joint spatial-spectral designs enforce mainlobe coverage~\cite{kang2019spatspect_taes}, sparse assignments leave the aggregate beam unchanged~\cite{wang2019sparse_taes}, and BiScatter supports two-way FMCW communication with sensing preserved~\cite{okubo2024biscatter_sigcomm}. Across these works, more payload implies more sensing perturbation, so compressing the payload beats upgrading the modulation. 

\textbf{Semantic communication.}
Task-oriented semantic communication compresses raw data into task-relevant representations so the channel carries only what the task needs. DeepSC~\cite{xie2021deepsc_tsp} and Deep JSCC~\cite{bourtsoulatze2019deepjscc_tccn} learn end-to-end semantic--channel mappings; the information bottleneck~\cite{tishby1999ib,shao2022ib_jsac} formalizes task-oriented compression, and VQ-VAE~\cite{vqvae2017nips} with robust digital bottlenecks~\cite{xie2023robust_jsac} makes representations exactly serializable. Joint sensing and task-oriented communication is emerging~\cite{sagduyu2024joint} but still assumes a conventional radio. No prior work couples semantic compression with sensing-preserving radar-centric constraints in communication-denied emergencies; we target this gap.


\section{System Design}
\label{sec:design}

\subsection{Overview}
\label{sec:design-overview}

\begin{figure*}[t]
\centering
\includegraphics[width=0.8\textwidth]{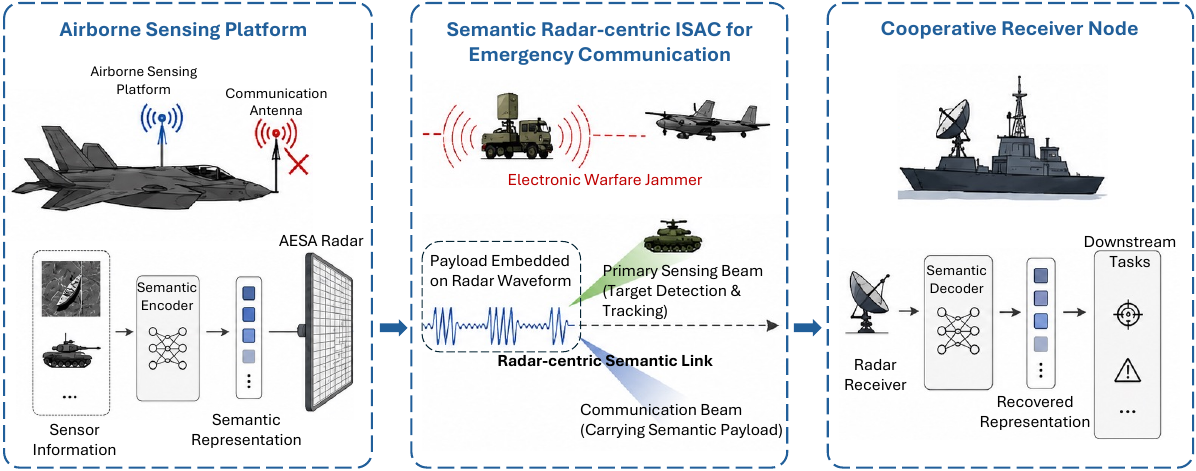}
\caption{Semantic radar-centric ISAC. A learned encoder compresses the application payload into a compact discrete semantic representation, which the radar transmits by selecting spatial-spectral configurations from a sensing-preserving subset; the receiver decodes the representation for downstream task inference.}
\label{fig:overview}
\vspace{-0.3cm}
\end{figure*}

We consider an emergency setting in which the dedicated radio is unavailable but a wideband phased-MIMO radar remains operational. The radar is sensing-critical, and communication is enabled only as a secondary fallback. The design must deliver a small mission-relevant message through the radar aperture while leaving sensing largely intact, and we solve the problem from two coupled directions. On the radar side, we restrict transmission to a conservative subset of spatial-spectral configurations, so communication is realized by selecting configurations rather than by injecting arbitrary symbols. On the semantic side, we learn a compact discrete representation that turns the application payload into a few tens to a few hundred bits, sized to fit within what the radar can safely carry. The two designs are coupled: the semantic bit budget is upper-bounded by the radar-safe rate, and the radar-safe rate is upper-bounded by a sensing-degradation budget. Fig.~\ref{fig:overview} summarizes the end-to-end workflow.

\subsection{System Model and Assumptions}
\label{sec:system-model}

We first fix the physical layer that both designs build on. The same radar emission is observed by two parties, the target (sensing) and a friendly listener (communication), so the two models below differ only in where the beam is steered and what we read off the return, not in the underlying waveform.

\subsubsection{Phased-MIMO Radar Platform}
We adopt a standard wideband near-field phased-MIMO radar built on a uniform planar array (UPA) with $M$ transmit subarrays, each an $N_x\times N_y$ UPA with inter-element spacing $d$; the element at indices $(n_x,n_y)$ of the $m$-th subarray sits at $[x_{m,n_x,n_y},\,y_{m,n_x,n_y}]^T$. The total bandwidth $W$ is split into $Q$ subcarriers $f_q = f_0 + (q-1)\tfrac{W}{Q}$, $q=1,\ldots,Q$. On each subcarrier the radar transmits $M$ orthogonal sequences of length $P$, $\mathbf{C}^q = [\mathbf{c}_1^q,\ldots,\mathbf{c}_M^q]^T$, satisfying $\tfrac{1}{P}\mathbf{C}^q(\mathbf{C}^q)^H = E_s\mathbf{I}_M$. The waveform of subarray $m$ on subcarrier $q$ at pulse $p$ and sample $k$ is $\psi_m^q(p,k) = c_m^q[p]\,e^{-j2\pi(q-1)(k-1)/Q}$.

\subsubsection{Sensing Model}
A point target at $[r_s\sin\theta_s,\,r_s\cos\theta_s]^T$ produces the intercepted signal
\begin{equation}
x_s(p,k)
=
\sum_{q=1}^{Q}\beta_q
\sum_{m=1}^{M}\mathbf{w}_m^{H}
\mathbf{a}_{m,q}(r_s,\theta_s)\,
\psi_m^q(p,k)
+ n_s(p,k),
\label{eq:sensing-rx}
\end{equation}
where $\beta_q$ is the propagation loss at subcarrier $q$, $\mathbf{w}_m$ is the subarray beamforming vector, and $\mathbf{a}_{m,q}(r_s,\theta_s)$ is the per-element spherical-wavefront near-field steering vector, retained because wideband near-field operation invalidates planar approximations. Intuitively, the beamformers $\{\mathbf{w}_m\}$ are the shared resource: the same vectors that focus energy on the target for sensing are reused, within limits, to illuminate the listener for communication.

\subsubsection{Communication Receiver Model}
We assume the emergency receiver is a friendly platform in the formation (e.g., a wingman) that already fields the same class of phased-MIMO radar. No extra hardware is needed: its existing radar receive chain recovers the payload via matched filtering and spectral-index detection, so the listen path is simply a radar receive mode rather than a dedicated communication terminal. We model it as a single-antenna listen point at known location $[r_c\sin\theta_c,\,r_c\cos\theta_c]^T$ with a dominant line-of-sight response, yielding
\begin{equation}
x_c(p,k)
=
\sum_{q=1}^{Q}\alpha_{ch,q}
\sum_{m=1}^{M}\mathbf{w}_m^{H}
\mathbf{a}_{m,q}(r_c,\theta_c)\,
\psi_m^q(p,k)
+ n_c(p,k),
\label{eq:comm-rx}
\end{equation}
with $\alpha_{ch,q}$ the known channel coefficient on subcarrier $q$ and $n_c(p,k)$ Gaussian noise.

\subsubsection{Emergency Semantic Payload}
The payload is not raw data but a learned discrete semantic representation $\mathbf{z}=E_\phi(\mathbf{x})$, where $E_\phi$ is the semantic encoder and $\mathbf{x}$ is the raw observation. The radar-centric layer delivers $\mathbf{z}$ in serialized form under the sensing-preservation constraint introduced next.

\subsection{Radar-Centric Communication Design}
\label{sec:radar-comm}

The central idea is to send bits not by modulating the waveform but by \emph{choosing} which radar configuration to transmit. Each admissible configuration is one symbol; we first enumerate the possible choices, then keep only those that leave sensing intact, and finally count how many bits the surviving choices can carry.

\subsubsection{Spatial-Spectral Configuration Codebook}
We adopt the spatial-spectral index-modulation principle, jointly indexing the spatial domain (subarray--waveform pairing) and the spectral domain (subcarrier selection steered toward the receiver via beam squint). A transmission configuration is the tuple $\Gamma_i = [m_i, s_i, q_i]$, where $m_i$ indexes the active transmit subarray, $s_i$ the orthogonal waveform assigned to it, and $q_i$ the selected subcarrier that steers energy toward the receiver; the full dictionary has cardinality $N_\Gamma = \binom{QM}{M}M!$. Communication is restricted to a conservative subset $\mathcal{G}_{\text{safe}}\subseteq\{\Gamma_i\}$ that satisfies the sensing-preservation constraint below.

\subsubsection{Sensing-Preserving Constraint}
A configuration is admitted into $\mathcal{G}_{\text{safe}}$ only if using it for communication does not visibly disturb the sensing picture. Concretely, the per-subarray beamformer $\mathbf{w}_m$ must (i) maintain mainlobe coverage over the sensing region across all subcarriers, (ii) bound sidelobe leakage, and (iii) deliver sufficient gain to the receiver at the selected subcarrier. The first two clauses protect sensing; the third guarantees the listener still hears the symbol. With sensing region $\Omega_s = \{(r,\theta):\,r_{\min}\!\le\!r\!\le\!r_{\max},\,\theta_{\min}\!\le\!\theta\!\le\!\theta_{\max}\}$ and receiver location $\Omega_c = \{(r_c,\theta_c)\}$, the sensing-preservation budget is $L_{\text{sense}}(\mathbf{W}) \le \epsilon$, instantiated as the beampattern-matching loss
$L_{\text{sense}}(\mathbf{W}) = \max_{(r_i,\theta_i)\in\Omega_s,\,q\in\mathcal{Q}}\big|\mathbf{w}_m^{H}\mathbf{a}_{m,q}(r_i,\theta_i)-e^{j\mu_q(r_i,\theta_i)}\big|$, with the sidelobe constraint $|\mathbf{w}_m^{H}\mathbf{a}_{m,q_m}(r_l,\theta_l)|\le\epsilon_{\text{side}}$ for $(r_l,\theta_l)\in\bar{\Omega}_s$ and the user-gain constraint $\mathbf{w}_m^{H}\mathbf{a}_{m,q_m}(r_c,\theta_c) = \zeta\,e^{j\tilde{\phi}_m}$, where $\mathcal{Q}$ is the set of selectable subcarriers and $\zeta$ is the required user-side gain. Configurations that violate any of these constraints are excluded from $\mathcal{G}_{\text{safe}}$.

\subsubsection{Representation-to-Configuration Mapping and Rate}
Let $\mathbf{b}_z = \text{Serialize}(\mathbf{z})$ be the serialized semantic bitstream. The radar-centric layer maps it to a sequence of configurations $\mathbf{b}_z \mapsto \{\Gamma^{(1)},\ldots,\Gamma^{(T)}\}$, each $\Gamma^{(t)}\in\mathcal{G}_{\text{safe}}$. The conservative radar rate is $R_{\text{radar}} = \lfloor \log_2 |\mathcal{G}_{\text{safe}}| \rfloor / T_{\text{sym}}$ with $T_{\text{sym}}=PT_{\text{pulse}}$, which is the operative ceiling; the unconstrained rate $\lfloor \log_2 N_\Gamma\rfloor/T_{\text{sym}}$ is recovered only when every configuration is sensing-safe. We deliberately trade a larger codebook for a smaller sensing-safe subset and offset the rate loss through semantic compression rather than stronger modulation.

\subsubsection{Radar Receiver Decoding}
Over a frame of $P$ pulses the receiver separates frequency components by a $Q$-point DFT and matched-filters with the orthogonal waveform set, $y_{q,m} = \tfrac{1}{P}\sum_{p=1}^{P} x^q[p]\,(c_m^q[p])^{H}$, then recovers $[\hat{m},\hat{s},\hat{q}]$ from the dominant spectral index and the associated phase structure. The decoded sequence is remapped to a bitstream $\hat{\mathbf{b}}_z$ and deserialized to $\hat{\mathbf{z}}$. When $\hat{\mathbf{b}}_z=\mathbf{b}_z$, the radar link is bit-transparent and any remaining end-to-end loss comes solely from semantic compression.

\subsection{Semantic Representation Learning Design}
\label{sec:semantic}

The radar side fixes a tight bit budget; the semantic side learns to say as much as the task needs within it. We therefore design the encoder so that its output is small, exactly serializable, and still sufficient for the downstream decision.

\subsubsection{Discrete Representation and Bit Budget}
The semantic encoder must produce a representation that is (i) compact, so its bit length fits the radar-safe budget; (ii) discrete, so it is exactly serializable and free from floating-point quantization noise; and (iii) task-sufficient, so the downstream decoder retains accuracy. We model the representation as a sequence of $L$ tokens $\mathbf{z} = [z_1,\ldots,z_L]$ with $z_i\in\{1,\ldots,K\}$, giving a total payload size $B_z = L\,\lceil \log_2 K\rceil$.

\subsubsection{Model Architecture}
The model has three components evaluated as $\mathbf{h} = E_\phi(\mathbf{x})$, $\mathbf{z} = Q(\mathbf{h})$, $\hat{y} = D_\theta(\mathbf{z})$. We instantiate $Q(\cdot)$ as a vector-quantized bottleneck: the encoder outputs $L$ continuous feature vectors $\mathbf{h} = [\mathbf{h}_1,\ldots,\mathbf{h}_L]$, and each $\mathbf{h}_i$ is mapped to the nearest entry of a learned codebook $\mathcal{V}=\{\mathbf{v}_1,\ldots,\mathbf{v}_K\}$ via $z_i = \arg\min_k\|\mathbf{h}_i-\mathbf{v}_k\|_2^2$. Only the indices $[z_1,\ldots,z_L]$ are transmitted; alternative bottlenecks (binary, learned tokenization) are compatible with the same design.

\subsubsection{Training Objective and Radar-Aware Constraint}
The model is trained end-to-end with $\mathcal{L} = \mathcal{L}_{\text{task}} + \lambda\,\mathcal{L}_{\text{rate}}$ and $\mathcal{L}_{\text{task}} = \ell\big(D_\theta(Q(E_\phi(\mathbf{x}))),\,y\big)$. When $(L,K)$ are fixed, the rate term degenerates into the design-time constraint $B_z \le B_{\max}$, with the radar-imposed budget $B_{\max} = R_{\text{radar}}\,T_{\max}$ and $T_{\max}$ the emergency latency budget. This couples the encoder directly to the radar-safe channel:
\begin{equation}
L\,\lceil \log_2 K\rceil \le R_{\text{radar}}\,T_{\max},
\label{eq:coupling}
\end{equation}
the bridge between semantic learning and radar communication. We compress not to save bandwidth but to bound sensing disturbance. In practice, given $R_{\text{radar}}$ and $T_{\max}$ we choose $(L,K)$ satisfying \eqref{eq:coupling}, train $E_\phi$, $Q$, and $D_\theta$ on paired samples $(\mathbf{x},y)$, and retain the smallest representation that meets the task-accuracy target.

\subsubsection{Joint Design Objective}
The complete system is a jointly constrained optimization coupling the semantic encoder/decoder, the discrete bottleneck, and the conservative embedding subset,
\begin{equation}
\begin{aligned}
\min_{E_\phi,Q,D_\theta}\;
& \mathbb{E}\big[\ell(D_\theta(\mathbf{z}),y)\big] \\
\text{s.t.}\;
& \mathbf{z}=Q(E_\phi(\mathbf{x})), \\
& B_z \le R_{\text{radar}}(\mathcal{G}_{\text{safe}})\,T_{\max}, \\
& L_{\text{sense}}(\mathcal{G}_{\text{safe}}) \le \epsilon ,
\end{aligned}
\label{eq:joint-obj}
\end{equation}
where $B_z$ is the semantic payload size, $R_{\text{radar}}(\mathcal{G}_{\text{safe}})$ the conservative radar rate, $T_{\max}$ the latency budget, and $L_{\text{sense}}\le\epsilon$ the sensing-preservation budget. Compared with throughput-maximizing radar-centric ISAC, our objective targets the smallest task-sufficient representation under a sensing-preservation constraint, turning the same spatial-spectral degrees of freedom into a conservative, radar-safe emergency channel.

\section{Evaluation}
\label{sec:evaluation}

We evaluate our framework along four research questions. \textbf{(RQ1)} Can the application payload be compressed to a small number of semantic bits without sacrificing task accuracy? \textbf{(RQ2)} Can the radar-centric link deliver these short payloads reliably across SNR? \textbf{(RQ3)} Does end-to-end task accuracy after radar transmission track the channel-free baseline? \textbf{(RQ4)} How much sensing degradation does the communication embedding introduce? We use two modalities, SAR target chips (MSTAR) and text classification (AG~News).

\begin{table*}[t]
\centering
\caption{Radar-centric link metrics on MSTAR and AG~News. Sem-64/Sem-128/Sem-256: 64/128/256-bit semantic payloads. BER/SER: bit and symbol error rates; ER: exact payload recovery (\%).}
\label{tab:radar-snr}
\scriptsize
\setlength{\tabcolsep}{2pt}
\renewcommand{\arraystretch}{1.05}
\begin{tabular*}{\textwidth}{@{\extracolsep{\fill}}llrrrrrrrrrrrr}
\toprule
\multirow{2}{*}{Data} & \multirow{2}{*}{SNR} & \multicolumn{3}{c}{Sem-64} & \multicolumn{3}{c}{Sem-128} & \multicolumn{3}{c}{Sem-256} & \multicolumn{3}{c}{Int8 latent} \\
\cmidrule(lr){3-5}\cmidrule(lr){6-8}\cmidrule(lr){9-11}\cmidrule(lr){12-14}
& & BER & SER & ER(\%) & BER & SER & ER(\%) & BER & SER & ER(\%) & BER & SER & ER(\%) \\
\midrule
\multirow{10}{*}{MSTAR}
& $-15$ & 0.4585 & 0.6955 & 0.00 & 0.4598 & 0.6864 & 0.00 & 0.4634 & 0.7075 & 0.00 & 0.4530 & 0.6645 & 0.00 \\
& $-10$ & 0.4319 & 0.6577 & 0.00 & 0.4235 & 0.6437 & 0.00 & 0.4350 & 0.6712 & 0.00 & 0.4181 & 0.6254 & 0.00 \\
& $-5$  & 0.3839 & 0.5953 & 0.00 & 0.3756 & 0.5793 & 0.00 & 0.3857 & 0.5998 & 0.00 & 0.3624 & 0.5583 & 0.00 \\
& $0$   & 0.3132 & 0.4882 & 0.00 & 0.3001 & 0.4671 & 0.00 & 0.3142 & 0.4855 & 0.00 & 0.2814 & 0.4435 & 0.00 \\
& $5$   & 0.2060 & 0.3056 & 0.00 & 0.1904 & 0.2896 & 0.00 & 0.2101 & 0.3079 & 0.00 & 0.1797 & 0.2769 & 0.00 \\
& $10$  & 0.0794 & 0.1105 & 1.50 & 0.0747 & 0.1050 & 0.00 & 0.0818 & 0.1118 & 0.00 & 0.0674 & 0.0976 & 0.00 \\
& $15$  & 0.0076 & 0.0091 & 74.75 & 0.0066 & 0.0086 & 56.50 & 0.0066 & 0.0083 & 35.50 & 0.0058 & 0.0075 & 0.00 \\
& $20$  & 0.0000 & 0.0000 & 100.00 & 0.0000 & 0.0000 & 100.00 & 0.0000 & 0.0000 & 99.50 & 0.0000 & 0.0000 & 99.25 \\
& $25$  & 0.0000 & 0.0000 & 100.00 & 0.0000 & 0.0000 & 100.00 & 0.0000 & 0.0000 & 100.00 & 0.0000 & 0.0000 & 100.00 \\
& $30$  & 0.0000 & 0.0000 & 100.00 & 0.0000 & 0.0000 & 100.00 & 0.0000 & 0.0000 & 100.00 & 0.0000 & 0.0000 & 100.00 \\
\midrule
\multirow{10}{*}{AG~News}
& $-15$ & 0.4650 & 0.7002 & 0.00 & 0.4579 & 0.6963 & 0.00 & 0.4604 & 0.7020 & 0.00 & 0.4567 & 0.6814 & 0.00 \\
& $-10$ & 0.4266 & 0.6522 & 0.00 & 0.4283 & 0.6579 & 0.00 & 0.4334 & 0.6649 & 0.00 & 0.4246 & 0.6426 & 0.00 \\
& $-5$  & 0.3846 & 0.5962 & 0.00 & 0.3785 & 0.5887 & 0.00 & 0.3824 & 0.5958 & 0.00 & 0.3702 & 0.5739 & 0.00 \\
& $0$   & 0.3104 & 0.4797 & 0.00 & 0.3060 & 0.4751 & 0.00 & 0.3120 & 0.4830 & 0.00 & 0.2930 & 0.4593 & 0.00 \\
& $5$   & 0.1950 & 0.2930 & 0.00 & 0.2013 & 0.2992 & 0.00 & 0.2048 & 0.3043 & 0.00 & 0.1911 & 0.2889 & 0.00 \\
& $10$  & 0.0753 & 0.1060 & 2.00 & 0.0782 & 0.1086 & 0.00 & 0.0784 & 0.1090 & 0.00 & 0.0725 & 0.1026 & 0.00 \\
& $15$  & 0.0068 & 0.0079 & 78.75 & 0.0067 & 0.0082 & 60.00 & 0.0062 & 0.0078 & 37.00 & 0.0063 & 0.0080 & 0.00 \\
& $20$  & 0.0000 & 0.0000 & 100.00 & 0.0000 & 0.0000 & 100.00 & 0.0000 & 0.0000 & 99.75 & 0.0000 & 0.0000 & 98.75 \\
& $25$  & 0.0000 & 0.0000 & 100.00 & 0.0000 & 0.0000 & 100.00 & 0.0000 & 0.0000 & 100.00 & 0.0000 & 0.0000 & 100.00 \\
& $30$  & 0.0000 & 0.0000 & 100.00 & 0.0000 & 0.0000 & 100.00 & 0.0000 & 0.0000 & 100.00 & 0.0000 & 0.0000 & 100.00 \\
\bottomrule
\end{tabular*}
\vspace{-0.3cm}
\end{table*}

\begin{table*}[t]
\centering
\caption{End-to-end task accuracy (\%) after radar transmission. Genie: reference accuracy with error-free payload delivery; other SNR columns: accuracy after radar delivery. ER@15: exact recovery (\%) at 15\,dB under the same end-to-end protocol.}
\label{tab:e2e-snr}
\scriptsize
\setlength{\tabcolsep}{2pt}
\renewcommand{\arraystretch}{1.05}
\begin{tabular*}{\textwidth}{@{\extracolsep{\fill}}llrrrrrrrrrr}
\toprule
\multirow{2}{*}{Data} & \multirow{2}{*}{Payload} &
\multirow{2}{*}{Genie} & \multicolumn{9}{c}{Receive SNR (dB)} \\
\cmidrule(lr){4-12}
& & & $-$15 & $-$10 & $-$5 & 0 & 5 & 10 & 15 & ER@15 & 20 \\
\midrule
\multirow{4}{*}{MSTAR}
& Sem-64  & 84.50 & 24.25 & 25.75 & 32.00 & 51.50 & 68.75 & 81.50 & 83.75 & 75.75 & 84.50 \\
& Sem-128 & 90.00 & 33.75 & 31.75 & 46.50 & 65.25 & 83.75 & 89.75 & 90.25 & 59.25 & 90.00 \\
& Sem-256 & 86.00 & 20.00 & 21.50 & 21.00 & 32.50 & 59.75 & 83.75 & 86.00 & 34.50 & 86.00 \\
& Int8    & 88.00 & 38.00 & 54.00 & 70.25 & 81.25 & 90.00 & 92.00 & 88.75 & 0.00  & 88.00 \\
\midrule
\multirow{4}{*}{AG~News}
& Sem-64  & 84.50 & 30.75 & 31.75 & 48.25 & 67.00 & 83.50 & 84.50 & 84.50 & 74.00 & 84.50 \\
& Sem-128 & 86.00 & 31.25 & 36.25 & 45.50 & 68.25 & 85.50 & 85.50 & 86.00 & 57.75 & 86.00 \\
& Sem-256 & 84.50 & 25.00 & 32.00 & 32.75 & 48.75 & 78.50 & 84.00 & 84.50 & 29.25 & 84.50 \\
& Int8    & 83.00 & 51.50 & 67.50 & 79.00 & 84.75 & 84.75 & 82.00 & 83.50 & 0.00  & 83.00 \\
\bottomrule
\end{tabular*}
\vspace{-0.3cm}
\end{table*}

\subsection{Experimental Setup}
\label{sec:eval-setup}

\textbf{Tasks and datasets.} On MSTAR we perform four-way SAR target classification (\emph{bmp2\_tank}, \emph{btr70\_transport}, \emph{t72\_tank}, \emph{SLICY}) on 3{,}549 chips with an 80/20 train/test split. On AG~News we perform four-way news classification on a stratified subset of 32{,}000 training and 4{,}000 test examples.

\textbf{Payload families.} The same five payloads are evaluated on both datasets: a raw-input proxy, an int8 continuous latent, and three discrete semantic representations of 64, 128, and 256 bits, denoted Sem-64, Sem-128, and Sem-256.

\textbf{Radar simulator.} A single physical layer is used throughout: carrier 10\,GHz, bandwidth 500\,MHz, $Q\!=\!64$ OFDM subcarriers, $P\!=\!32$ pulses per symbol, a $32\!\times\!32$ UPA partitioned into $4\!\times\!4$ subarrays, and four selected subcarriers, with $T_{\mathrm{pulse}}=1.28\!\times\!10^{-7}$\,s and $T_{\mathrm{sym}}=PT_{\mathrm{pulse}}$. After sensing-preserving filtering, four spatial-spectral configurations remain for communication, giving 2\,bits per radar symbol. The SNR sweep covers $-15$\,dB to $30$\,dB in 5\,dB steps, with 400 Monte~Carlo payload transmissions per point. Table~\ref{tab:radar-snr} summarizes the radar-centric sweep; the per-payload time-on-air is in Table~\ref{tab:latency}.

\textbf{End-to-end protocol.} Each test sample is live-encoded into a bitstream, transmitted through the same simulator, recovered by maximum-likelihood spectral-index detection, and classified by the semantic decoder alone, without re-running the upstream encoder.

\begin{table}[t]
\centering
\caption{Radar time-on-air under frame-based transmission ($T_{\mathrm{pulse}}\!=\!1.28\!\times\!10^{-7}$\,s, $T_{\mathrm{sym}}\!=\!PT_{\mathrm{pulse}}$, $T_{\mathrm{payload}}\!=\!N_{\mathrm{sym}}T_{\mathrm{sym}}$).}
\label{tab:latency}
\scriptsize
\setlength{\tabcolsep}{5pt}
\begin{tabular}{lrrr}
\toprule
Payload & Bits & Sym. & Time (ms) \\
\midrule
Sem-64     & 64   & 32   & 0.131 \\
Sem-128    & 128  & 64   & 0.262 \\
Sem-256    & 256  & 128  & 0.524 \\
Int8 latent & 2048 & 1024 & 4.194 \\
\bottomrule
\end{tabular}
\vspace{-0.3cm}
\end{table}

\subsection{Semantic Compression Efficiency (RQ1)}
\label{sec:eval-rq1}

\begin{table}[t]
\centering
\caption{Semantic compression efficiency. CR is relative to the raw-input proxy; Acc. is classification accuracy (\%).}
\label{tab:semantic-efficiency}
\scriptsize
\setlength{\tabcolsep}{3.5pt}
\begin{tabular}{llrrr}
\toprule
Data & Payload & Bits & CR & Acc. (\%) \\
\midrule
\multirow{5}{*}{MSTAR}
& Int8 latent & 2048   & 1$\times$   & 90.28 \\
& Sem-64      & 64     & 32$\times$ & 85.07 \\
& Sem-128     & 128    & 16$\times$ & 87.46 \\
& Sem-256     & 256    & 8$\times$  & 79.86 \\
\midrule
\multirow{5}{*}{AG~News}
& Int8 latent & 2048 & 1$\times$ & 87.78 \\
& Sem-64      & 64   & 32$\times$   & 87.93 \\
& Sem-128     & 128  & 16$\times$   & 88.30 \\
& Sem-256     & 256  & 8$\times$   & 88.03 \\
\bottomrule
\end{tabular}
\vspace{-0.35cm}
\end{table}

Table~\ref{tab:semantic-efficiency} reports task accuracy as a function of payload size. On MSTAR, the int8 latent attains the highest accuracy of any representation. The discrete representations push compression by several of magnitude with only marginal accuracy loss; the 128-bit variant comes closest to the int8 variant, while the largest one underperforms its smaller-codebook siblings, which we attribute to the difficulty of training a large codebook on a small SAR dataset rather than to any limitation of the radar layer. On AG~News every learned representation matches or slightly exceeds the int8 latent while occupying a small fraction of its size. The pattern across modalities supports the core claim: when the receiver only needs the task answer rather than a faithful reconstruction, tens to a few hundred bits suffice.

\subsection{Radar-Centric Semantic Transmission (RQ2)}
\label{sec:eval-rq2}

Table~\ref{tab:radar-snr} summarizes the full SNR sweep on both datasets. BER and SER fall sharply as SNR rises, and exact payload recovery (ER) transitions from near-zero at low SNR to near-perfect in the high-SNR regime on both datasets. Shorter payloads enter the useful operating regime earlier, because exact recovery requires every bit to be correct and therefore penalizes long payloads at borderline SNR. The link is essentially error-free at the top of the sweep across all payloads on both datasets, so in the moderate-to-high SNR regime the binding constraint is payload length rather than channel quality. When considered alongside Table~\ref{tab:latency}, these findings demonstrate precisely where semantic compression yields operational gains: short semantic packets reach reliable recovery at lower SNR and occupy a small fraction of the int8 latent's radar time-on-air, leaving the rest of the radar dwell for sensing.

\subsection{End-to-End Task Accuracy After Radar Transmission (RQ3)}
\label{sec:eval-rq2b}

Link-layer BER and exact recovery alone do not tell us whether the mission task survives delivery. Table~\ref{tab:e2e-snr} reports the genie baseline ($\hat{\mathbf{b}}_z=\mathbf{b}_z$) and end-to-end task accuracy after radar delivery (channel-free accuracies are in Table~\ref{tab:semantic-efficiency}). On both modalities, accuracy collapses toward chance at low SNR, rises sharply past the moderate regime, and matches the genie baseline by 20\,dB (25--30\,dB are identical and omitted). At 15\,dB, end-to-end accuracy already nears the genie while exact recovery remains far lower, because discrete-token errors often fall in task-equivalent codebook regions.

\subsection{Sensing Preservation Under Communication Embedding (RQ4)}
\label{sec:eval-rq3}

\begin{table}[t]
\centering
\caption{Sensing preservation versus embedding strength $\eta$. Metrics are dataset-independent; payload-specific time-on-air is in Table~\ref{tab:latency}.}
\label{tab:sensing-preservation}
\scriptsize
\setlength{\tabcolsep}{3.5pt}
\begin{tabular}{lrrrr}
\toprule
Mode & $\eta$ & Tgt. loss (dB) & Side inc. (dB) & User gain (dB) \\
\midrule
Sense-only & 0.00 & 0.0000  & 0.0000 & 33.56 \\
Conserv.   & 0.03 & $-0.0020$ & 0.0134 & 35.89 \\
Aggress.   & 0.15 & $-0.0665$ & 0.0314 & 42.35 \\
\bottomrule
\end{tabular}
\vspace{-0.3cm}
\end{table}

Table~\ref{tab:sensing-preservation} reports the sensing-side metrics for two communication-embedding settings against a sensing-only baseline. The sensing metrics depend only on the radar geometry and the embedding strength, so they are dataset-independent; the per-dataset payload only affects time-on-air (Table~\ref{tab:latency}). At the conservative operating point, the mainlobe loss toward the target is small and the sidelobe perturbation is bounded to a fraction of a decibel, while the user-side gain rises noticeably above the sensing-only reference, enough to support the short semantic payloads of Table~\ref{tab:radar-snr}. The aggressive setting trades additional user gain for measurable mainlobe loss and is not used in the reported link experiments. The conservative configuration is therefore the natural default emergency mode: it provides a usable communication path for compact semantic payloads while keeping the radar beampattern almost unchanged, exactly the operating regime that the joint design objective targets.

\section{Conclusion}
\label{sec:conclusion}

We investigated emergency communication on military platforms through a radar-centric ISAC link that preserves the radar’s primary sensing function. Our framework combines a vector-quantized semantic encoder with a wideband phased-MIMO radar operating within a conservative spatial-spectral subset, matching the semantic bit budget to the radar-safe rate under a bounded sensing-loss constraint. Experiments on SAR target and text classification show that compact semantic payloads retain high end-to-end accuracy with negligible beampattern distortion. These results demonstrate that sensing-critical radar can provide a practical emergency link without compromising its primary mission. Future work will evaluate mission-specific waveforms, hardware impairments, and field deployment.

\section*{Acknowledgments} 
This work was supported in part by the Office of Naval Research under grant N00014-24-1-2730, the National Science Foundation under grants 2433904, 2312447, 2247560, and 2235232, and the Virginia Commonwealth Cyber Initiative.

\bibliographystyle{IEEEtran}
\bibliography{reference}

\end{document}